\documentclass{article}
\usepackage{spconf,amsmath,graphicx}
\usepackage{amsfonts} 
\usepackage{subfigure}

\usepackage{cite}
\usepackage{multirow}
\usepackage{threeparttable}
\usepackage{booktabs}
\usepackage{tabularx}
\newcolumntype{C}[1]{>{\centering\arraybackslash}p{#1}}
\newcolumntype{L}[1]{>{\raggedright\arraybackslash}p{#1}}
\newcolumntype{R}[1]{>{\raggedleft\arraybackslash}p{#1}}

\title{Music Source Separation with Deep Equilibrium Models}
\name{Yuichiro Koyama$^{\star}$,
\quad Naoki Murata$^{\star}$,
\quad Stefan Uhlich$^{\dagger}$,
\quad Giorgio Fabbro$^{\dagger}$,}
\secondlinename{Shusuke Takahashi$^{\star}$,
\quad Yuki Mitsufuji$^{\star}$}

\address{$^{\star}$ Sony Group Corporation, Tokyo, Japan \qquad\qquad
      $^{\dagger}$Sony Europe B.V., Stuttgart, Germany}

\begin{document}
\ninept

\maketitle

\begin{abstract}
While deep neural network-based music source separation (MSS) is very effective and achieves high performance, its model size is often a problem for practical deployment. Deep implicit architectures such as deep equilibrium models (DEQ) were recently proposed, which can achieve higher performance than their explicit counterparts with limited depth while keeping the number of parameters small. This makes DEQ also attractive for MSS, especially as it was originally applied to sequential modeling tasks in natural language processing and thus should in principle be also suited for MSS. However, an investigation of a good architecture and training scheme for MSS with DEQ is needed as the characteristics of acoustic signals are different from those of natural language data. Hence, in this paper we propose an architecture and training scheme for MSS with DEQ. Starting with the architecture of Open-Unmix (UMX), we replace its sequence model with DEQ. We refer to our proposed method as DEQ-based UMX (DEQ-UMX). Experimental results show that DEQ-UMX performs better than the original UMX while reducing its number of parameters by 30\%.
\end{abstract}
\section{Introduction}
\label{sec:intro}
Deep neural network~(DNN)-based approaches are highly effective in achieving high performance in acoustic signal processing tasks such as music source separation (MSS)~\cite{stefan2015deep,stefan2017improving,naoya2017multi, stoter2019open,defossez2019music,choi2020investigating,takahashi2021densely,xumx,mitsufuji2021music,kong2021decoupling}.
However, their model size is often a problem for practical deployment.
For example, downloading a software package including larger DNN models can take a considerable amount of time and frustrate users. 
Also, larger DNN models require more random access memory and read-only memory, which could be a problem when the available on-device memory size is limited.
Therefore, methods for model size reduction which preserve the performance of the original network are worth exploring.

Deep implicit layers~\cite{amos2017optnet,chen2018neural,bai2019equilibrium,bai2020multiscale,fung2021fixed,wang2021approximate} define layers implicitly such that 
input and output have to satisfy some joint conditions, e.g., that they are the equilibrium points of an equation:
they were recently proposed in the field of machine learning and achieved higher performance than typical networks whose architecture is explicitly defined.
Deep equilibrium models (DEQ)~\cite{bai2019equilibrium} are an instance of deep implicit approaches, and their layer output is calculated by finding a root of an equation (i.e., equilibrium point) parameterized by the layer input.
DEQ is also considered to be a weight-tied network~\cite{bai2018trellis} that has approximately infinite layers.
It was shown that DEQ is especially suited for sequential modeling tasks and outperformed other existing methods 
with a number of parameters comparable to that of only a few explicit layers.

For example, DEQ-TrellisNet and DEQ-Transformer achieved state-of-the-art performance in large-scale language modeling tasks using WikiText-103~\cite{bai2019equilibrium}.
Multi-scale DEQ, which computes several equilibrium points at different scales to deal with high-resolution images, also outperformed other existing methods in some image-processing tasks such as ImageNet classification and semantic segmentation~\cite{bai2020multiscale}. 
However, such DEQ-based approaches have not been explored in audio source separation.
In \cite{Takeuchi2020real}, equilibriated recurrent neural networks (ERNN)~\cite{kag2019rnns} were applied to speech enhancement.
ERNN is a similar approach to DEQ as it
approximately estimates equilibrium points of ordinary differential equations using the implicit Euler method.
However, its motivation is to avoid gradient problems and stabilize the training procedure.
Also, ERNN replaces one recurrent layer with ERNN layers whereas DEQ replaces any repeating layers with a DEQ-based layer.

Audio source separation requires sequence modeling as acoustic signals have a sequential structure.
A network architecture for source separation is typically composed of an encoder block, a sequence modeling block, and a decoder block~\cite{Pariente2020Asteroid}.
The sequence modeling block is usually implemented with recurrent layers or 1-D or 2-D convolutional layers.
For instance, Open-Unmix (UMX)~\cite{stoter2019open} employs a long short-term memory (LSTM) network, while Conv-TasNet~\cite{luo2019conv} employs a temporal convolutional network (TCN).
Therefore, DEQ is expected to be also suited as sequence modeling block for a source separation architecture and to contribute to a performance improvement, together with a reduction in the number of parameters.
However, since the characteristics of acoustic signals
are different from those of natural language data, an architecture and training scheme appropriate for source separation need to be designed.

In this paper, we propose to apply DEQ to MSS. 
Using the architecture of Open-Unmix (UMX)~\cite{stoter2019open} as a starting point, 
we replace its sequence model with DEQ.
We refer to our proposed method as DEQ-based UMX (DEQ-UMX).
Usually DEQ-based approaches tend to require a high computational cost due to the number of iterations needed to find the equilibrium point and obtain a sufficiently large receptive field.
Nevertheless, we expect that DEQ-UMX does not require a too high computational cost and can achieve sufficient performance with a few iterations by employing bidirectional LSTM (BLSTM) as a core function for DEQ.
This is because the BLSTM layer will utilize all information of the input sequence even if the number of iterations is small. 

As mentioned earlier, DEQ can be interpreted as an infinite stack of weight-tied networks, therefore we start by experimenting with a weight-tied network for MSS.
After confirming its effectiveness, we move to DEQ for further performance improvement.
Experimental results show that DEQ-UMX performs better than the original UMX while reducing the number of parameters.

\begin{figure*}[htb]
  \centering
  \includegraphics[width=12.0cm]{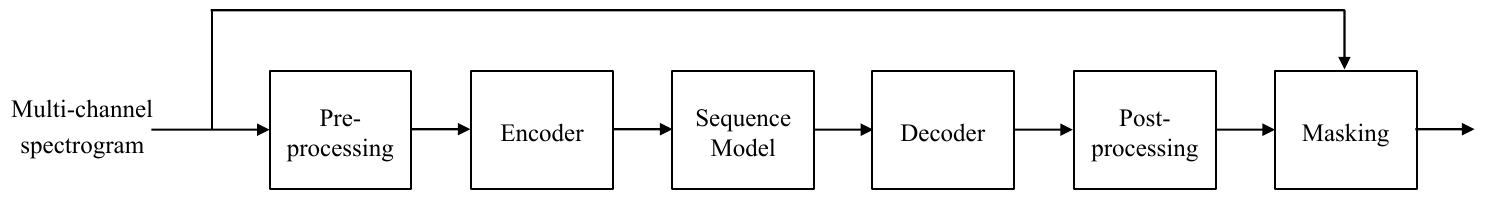} 
  \vspace{-2mm}
  \caption{General network architecture for STFT-based source separation.}
\label{fig:general_pipeline}
\vspace{-3mm}
\end{figure*}
\begin{figure}[htb]
  \centering
  \includegraphics[width=7.0cm]{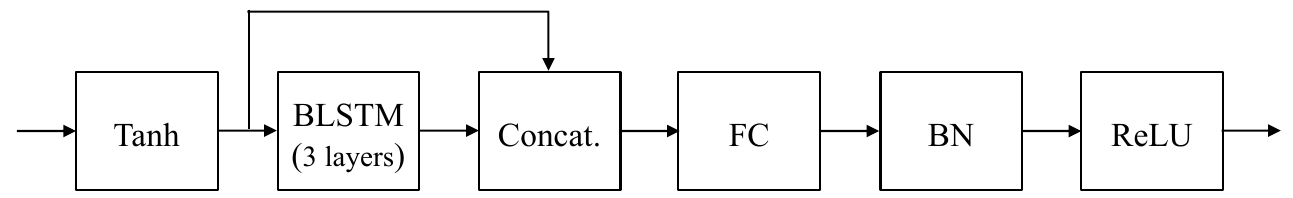} 
  \vspace{-2mm}
  \caption{Sequence model of UMX}
\label{fig:umx_sep}
\vspace{-3mm}
\end{figure}
\section{Related Work}
\label{sec:related_work}

In this section we review weight-tied networks and DEQ to support our contributions, and Open-Unmix to provide information about a general architecture for DNN-based source separation.
\subsection{Weight-tied network}
A weight-tied network is realized by employing the same transformation $f_\theta$, parameterized by $\theta$, in each layer of a typical skip-connection-based architecture as
\begin{equation}
\mathbf{z}^{[i+1]}_{1:T} = f_\theta\left(\mathbf{z}^{[i]}_{1:T}; \mathbf{x}_{1:T} \right) ,
\label{eq:wt_net}
\end{equation}
where $\mathbf{x}_{1:T} \in \mathbb{R}^{T\times p}$ is the input sequence and $\mathbf{z}^{[i]}_{1:T}\in \mathbb{R}^{T\times q}$ is the hidden sequence in the $i$-th layer~\cite{bai2018trellis}.
Such transformation can also be interpreted as a sequence of identical transformations, where each iteration is identified by the index $i$.
This sequence is typically terminated after a fixed number of iterations $L$ (i.e., $i = 0,1,\dots,L-1$).
The number of iterations is equivalent to the number of function evaluations (NFE), thus $L$ is also called NFE.
TrellisNet~\cite{bai2018trellis}, which is implemented by using TCN as $f_\theta$, outperformed other typical DNN-based approaches on some natural language processing (NLP) tasks without a noticeable increase of the number of parameters.

\subsection{Deep Equilibrium Model}
In \cite{bai2019equilibrium}, it was empirically shown that \eqref{eq:wt_net} tends to converge to an equilibrium point $\mathbf{z}^{\star}_{1:T}$ as the layer index $i$ increases.
The equilibrium point $\mathbf{z}^{\star}_{1:T}$ satisfies the following condition: 
\begin{equation}
\mathbf{z}^{\star}_{1:T} = f_\theta\left(\mathbf{z}^{\star}_{1:T}; \mathbf{x}_{1:T} \right),
\label{eq:f_theta}
\end{equation}
and the output of DEQ is the equilibrium point itself. DEQ showed further improvement over weight-tied networks on some NLP tasks~\cite{bai2019equilibrium}.

The forward pass of DEQ is performed by finding the equilibrium point $\mathbf{z}^{\star}_{1:T}$.
Some root-finding solvers for nonlinear equations such as Newton's method or quasi-Newton methods can be utilized to find the equilibrium point as follows:
\begin{equation}
\mathbf{z}^{[i+1]}_{1:T} = 
\mathbf{z}^{[i]}_{1:T} - \alpha B g_{\theta} \left(\mathbf{z}^{[i]}_{1:T};\mathbf{x}_{1:T} \right)
\quad (i=0,1,\dots,L_{\text{stop}}-1),
\label{eq:newton}
\end{equation}
where $g_{\theta}(\mathbf{z}_{1:T}^{[i]}; \mathbf{x}_{1:T}) = f_{\theta}(\mathbf{z}^{[i]}_{1:T}; \mathbf{x}_{1:T}) - \mathbf{z}^{[i]}_{1:T}$, 
$B$ is the inverse Jacobian of $g_{\theta}$ (or its low-rank approximation) evaluated at $\mathbf{z}_{1:T}^{[i]}$, which is obtained by a root-finding solver,  $\alpha$ is the step size, and $L_{\text{stop}}$ is the NFE when the iteration is stopped.
The iteration stops if the norm of $g_{\theta}$ becomes smaller than a tolerance $\epsilon$ or the maximum number of function evaluations $L_{\text{max}}$ is reached. 
Broyden's method~\cite{broyden1965class}, which is a quasi-Newton method, is often used for DEQ as it computes the inverse Jacobian efficiently.

In DEQ, the backpropagation is performed on the basis of implicit differentiation~\cite{bai2019equilibrium}, which doesn't need to store the intermediate values of the solver.
Let $l$ be the loss function for optimizing the parameter $\theta$.
The derivative of $l$ with respect to $\theta$ can be calculated based on implicit differentiation as follows:
\begin{equation}
\frac{\partial l}{\partial \theta} = 
-\frac{\partial l}{\partial \mathbf{z}^{\star}_{1:T}}
\left( J^{-1}_{g_{\theta}}|_{\mathbf{z}^{\star}_{1:T}}\right)
\frac{\partial f_\theta\left(\mathbf{z}^{\star}_{1:T}; \mathbf{x}_{1:T} \right)}{\partial \theta},
\label{eq:backward}
\end{equation}
where $J^{-1}_{g_{\theta}}$ is the inverse Jacobian of $g_{\theta}$ evaluated at the equilibrium point $\mathbf{z}_{1:T}^{\star}$, which appears in the implicit differentiation theorem. 
The operation $-\frac{\partial l}{\partial \mathbf{z}^{\star}_{1:T}}
\left( J^{-1}_{g_{\theta}}|_{\mathbf{z}^{\star}_{1:T}}\right)$ can be performed by solving the equation in terms of $\mathbf{y}$: 
\begin{equation}
\left( J^{\top}_{g_{\theta}}|_{\mathbf{z}^{\star}_{1:T}}\right)\mathbf{y}^{\top} + 
\left(\frac{\partial l}{\partial \mathbf{z}^{\star}_{1:T}}\right)^{\top}
= \mathbf{0}
\label{eq:backward_eq}
\end{equation}
with solvers such as Broyden's method.

Jacobian-Free backpropagation (JFB)~\cite{fung2021fixed} was proposed as a more efficient technique for gradient calculation in DEQ. This method approximates the inverse Jacobian in \eqref{eq:backward} with an identity matrix, avoiding the Jacobian-based computation from \eqref{eq:backward_eq} and reducing computations. 
In \cite{fung2021fixed} it was shown that the gradient obtained by JFB is valid (i.e., a descent direction for the loss function) under certain conditions. Furthermore, since this method does not impose a condition on how close the solver's output is to the equilibrium point, it is expected to be robust to estimation errors by the solver.

\begin{figure*}[t]
\centering\subfigure[Proposed sequence model]
{\includegraphics[width=5.0cm]{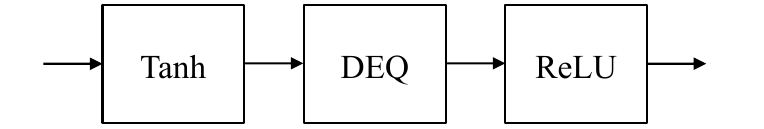} 
 \label{fig:deq_seq_model}}
\hspace{0mm}
\centering\subfigure[Function $f_{\theta}$ of the DEQ block]
{\includegraphics[width=11.0cm]{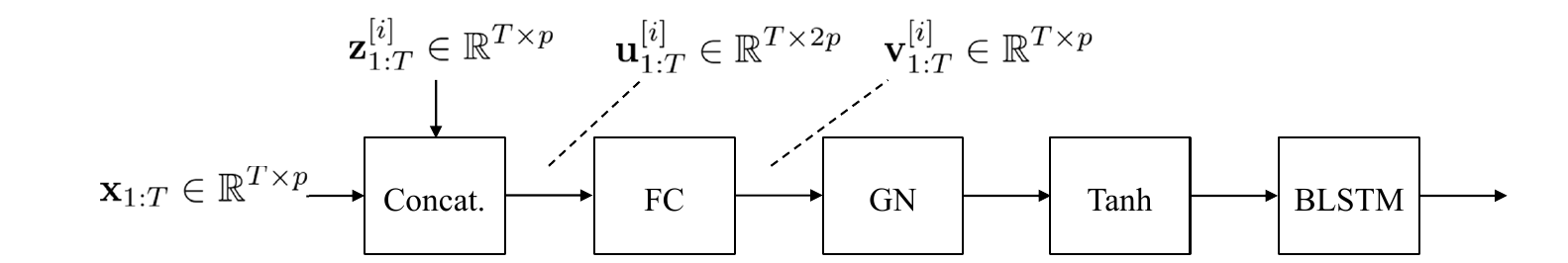} 
 \label{fig:deq_function_f}}
\hspace{0mm}
\vspace{-3mm}
\caption{The overview of the sequence model of the proposed method. We propose to replace the sequence model of the UMX with the DEQ-based sequence model.}
\label{fig:proposed_method}
\vspace{-3mm}
\end{figure*}

\subsection{Open-Unmix}

Fig.~\ref{fig:general_pipeline} is an example of a general network architecture for DNN-based audio source separation that receives a multi-channel (typically stereo in MSS) spectrogram obtained by a short-time Fourier transform (STFT) and outputs the spectrogram for the target audio source.
UMX is a typical example of such architecture, which shows performance close to state-of-the-art~\cite{stoter2019open}.

In the pre-processing block of UMX, the input spectrogram is first cropped to a maximum frequency (typically 16~kHz). Then the cropped spectrogram is fed into frame-wise linear transformations whose parameters are set to normalize the spectrogram to have zero mean and unitary standard deviation. The encoder block is composed of a fully-connected (FC) layer with batch normalization (BN).
The sequence modeling block is a key part of source separation as it deals with contextual information necessary to separate the mixed sources.
Fig.~\ref{fig:umx_sep} shows in detail the sequence model of UMX. The output of the encoder block is first normalized with a hyperbolic tangent function such that its range is between $-1$ and $1$, then fed into 3 consecutive BLSTM layers.
The normalized output of the encoder block and the output of the BLSTM block are concatenated and sequentially fed into another FC layer with BN and rectified linear unit (ReLU) activation.
The decoder block is composed of a FC layer with BN.
The post-processing block is a linear transformation whose parameters are initialized to 1.
The output of the post-processing block is used as a multiplicative mask on the input of the pre-processing module.
Such architecture is trained to separate a specific instrument.
Finally, the separated signals for each target instrument are refined by applying a multi-channel Wiener filter~(MWF)~\cite{nugraha2016multichannel,stefan2017improving} to the output of each network.

\vspace{-1mm}
\section{Proposed method}
\label{sec:proposed_method}
Although the DEQ already showed excellent performance in some NLP tasks, 
the characteristics of acoustic signals is different from that of natural language data.
Therefore, we devise in this section a suitable architecture and a training scheme for DEQ-based source separation.

\subsection{Forward pass}
\label{subsec:model}
We replace the sequence model of UMX with a DEQ-based sequence model while keeping the other blocks as they are.
Fig.~\ref{fig:proposed_method} describes the overview of the sequence model of the proposed method.
In the sequence model, the input sequence normalized by the hyperbolic tangent function is fed into the DEQ block followed by a ReLU as shown in Fig.~\ref{fig:deq_seq_model}.
Fig.~\ref{fig:deq_function_f} shows our proposed architecture for the function $f_{\theta}$ of the DEQ block, which is inspired by the original UMX.
Specifically, we adopt a concatenation block to merge the output of the BLSTM and the input sequence.
We define the size of hidden sequence $q$ as equivalent to that of the input sequence $p$ since the original UMX has the same size of output with the input.
Then the $i$-th concatenated sequence $\mathbf{u}^{[i]}_{1:T} \in \mathbb{R}^{T\times 2p}$ is fed into a FC layer such that its dimension is the same as the original input of the DEQ block, 
\begin{equation}
\mathbf{v}^{[i]}_{1:T} = \text{FC}(\mathbf{u}^{[i]}_{1:T}), 
\label{eq:fc}
\end{equation}
where $\mathbf{v}^{[i]}_{1:T} \in \mathbb{R}^{T\times p}$ is the $i$-th output sequence of the FC block.
Then the output sequence $\mathbf{v}^{[i]}_{1:T}$ is normalized with a group normalization (GN) block whose number of groups is one and fed into a hyperbolic tangent function followed by a BLSTM block with one layer as follows,
\begin{equation}
f_{\theta}\left(\mathbf{z}^{[i]}_{1:T};\mathbf{x}_{1:T} \right) = \text{BLSTM}(\tanh(\text{GN}(\mathbf{v}^{[i]}_{1:T}))).
\label{eq:BLSTM}
\end{equation}
The choice of GN is a deviation from the original UMX, which uses BN instead.
This modification is inspired by~\cite{bai2020multiscale} as BN often causes a poor conditioning of the Jacobian matrix.
We use Broyden's method to compute Eq.~\eqref{eq:newton} in the forward pass and obtain the hidden sequence $\mathbf{z}^{[i+1]}_{1:T}$.
Note that the tolerance $\epsilon$ and the maximum number of function evaluations $L_{\text{max}}$ during inference need to be the same with those of training.

Aside from the architecture shown in Fig.~\ref{fig:deq_function_f}, we preliminary evaluated two different architectures, where we substitute concatenation with weight-averaging or with masking.
We found that the architecture with the concatenation block achieved the best performance, which agrees with the tendency highlighted by \cite{braun2021towards}.
Therefore, we adopt the architecture shown in Fig.~\ref{fig:deq_function_f}.

To provide further insights into our architecture, an analogy with guided source separation (GSS)~\cite{kanda2019guided,speakerbeam2019,spex2020} may be helpful.
In the first iteration of the solver (i.e., $i=0$), $\mathbf{z}^{[0]}_{1:T}$ (which is an all-zero sequence) is fed into the function $f_{\theta}$ and concatenated with the input $\mathbf{x}_{1:T}$.
This process can be interpreted as if the function $f_{\theta}$ attempts to extract the desired sequence without any guide information.
In the first few iteration~(e.g., $i=0,1,2$), 
although the calculated $\mathbf{z}^{[i+1]}_{1:T}$ is still insufficient for the separation performance, it is used as the updated guide information for the next iteration.
The guide information gradually becomes more accurate with each iteration and finally a sufficiently informative $\mathbf{z}^{[L_{\text{stop}}]}_{1:T}$ can be obtained.

Using BLSTM in the function $f_\theta$ is a key point.
If a convolutional neural network (CNN)-based architecture such as TrellisNet~\cite{bai2018trellis} was assigned to $f_\theta$, it would require many NFE to obtain a sufficiently large receptive field. 
In fact, TrellisNet requires hundreds of iterations to converge, and dozens of iterations are performed in practice~\cite{bai2018trellis,bai2019equilibrium}.   
In contrast, we can expect that using BLSTM in the function $f_\theta$ enables us to utilize all information of the input sequence $\mathbf{x}_{1:T}$ even if the NFE is small.

\subsection{Backward pass and training}
Although a small NFE
contributes to reducing the computational cost, 
it could be a problem when using implicit differentiation for the backward pass because it is not guaranteed that the obtained hidden sequence satisfies the equilibrium condition sufficiently.
Therefore, since JFB is expected to be more robust to the estimation error of the solver, we use JFB for an efficient gradient calculation instead of using implicit differentiation-based backpropagation.

We use the same training setup (i.e., loss function, learning rate scheduler, etc...) as in the original UMX.
Following \cite{bai2019equilibrium},
the parameters of the DEQ-UMX are pre-trained as weight-tied network using the typical backpropagation with a pre-defined number of epochs for the stability of the training.
NFE of the weight-tied network and the number of epochs for pre-training are determined on the basis of the validation loss.

\vspace{-2mm}
\section{Experiment}
\begin{table*}
    \caption{Evaluation results on the MUSDB18 dataset. DEQ-UMX achieved the best average SDR while reducing the number of parameters.}

  \label{tab:result}
  \centering
 \scalebox{0.88}{ 
 \begin{tabular}{l|c|c|ccccc}
    \toprule

        \multirow{2}{*}{Method} &
        \multirow{2}{*}{\# Param. (M)} &
        \multirow{2}{*}{MACs (G) / 6sec.} & 
        \multicolumn{5}{c}{SDR [dB]}  
        
        \\ 
        &  &  & Vocals & Drums & Bass & Other & Avg.
        \\ 
        \midrule
        Open-Unmix (UMX)~\cite{stoter2019open} & 35.55 & 9.08 & 
        6.32 & 
        5.73 &
        \textbf{5.23} &
        4.02 &
        5.33 \\
        \midrule
        UMX large (4 layers in BLSTM) & 41.85 & 10.69 & 
        6.41 &
        5.94 &
        4.87 &
        \textbf{4.21} & 
        5.36 \\
        UMX large (5 layers in BLSTM) & 48.16 & 12.30 & 
        6.22 &
        5.75 &
        5.16 & 
        4.03 &
        5.29 \\ 
        UMX small & 25.15 & 6.42 & 
        6.15 &
        5.78 & 
        4.87 & 
        4.16 & 
        5.24 \\
        \midrule
        WT-UMX ($L=4$) & 25.06 & 12.29 &
        6.37 &
        6.12 & 
        5.20 & 
        3.94 &
        5.41 \\
        \midrule
        DEQ-UMX (with implicit diff.) & 25.06 & 18.74 &
        6.32 &
        5.94 & 
        5.05 & 
        4.07 & 
        5.34 \\
        \textbf{DEQ-UMX (with JFB, proposed)} & 25.06 & 18.74 &
        \textbf{6.60} &
        \textbf{6.17} & 
        5.14 & 
        4.20 & 
        \textbf{5.53} \\
        \bottomrule
    \end{tabular}
}
\vspace{-4mm}
\end{table*}
\subsection{Experimental settings}

We use the MUSDB18 dataset, which consists of 150 professionally recorded songs~\cite{rafii2017musdb18}.
For each song, the clean waveform of \emph{vocals}, \emph{drums}, \emph{bass}, and \emph{other} together with their mixture are available. Each audio track is stereo with a sampling frequency of 44.1~kHz.
Following the official split, we used 86~songs for the training, 14~songs for the validation, and 50~songs for the test set.
All networks in this experiment were trained on 6 seconds long segments using the Adam~\cite{kingma2015adam} optimizer and a weight decay of~$10^{-5}$.
The learning rate was initially set to 0.001 and decreased by 70\% if the average of the loss function on the validation set did not improve in 80 consecutive epochs.
The training was interrupted when the average of the loss function on the validation set did not improve in 300 consecutive epochs.
For inference, the full-length signals were processed with the trained networks.
MWF was also applied to the full-length signals.
Signal-to-distortion ratio (SDR)~\cite{vincent2006performance} was used for the evaluation.
We compute the SDR by taking the median over all frames of a song and then calculating the median over all songs, which is the standard way of SDR computation on MUSDB18.

\vspace{-2mm}
\subsection{Preliminary study: WT-UMX}
\begin{figure}[tb]
  \centering
  \includegraphics[width=6.4cm]{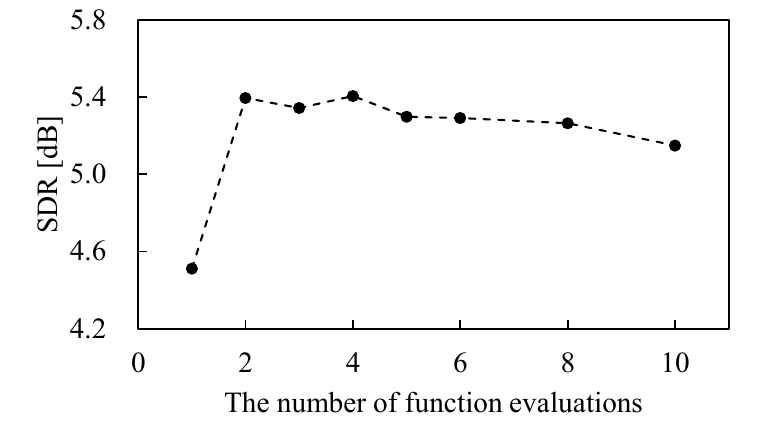}
 \vspace{-3mm}
 \caption{Relationship between NFE of WT-UMX and average SDR.}
\label{fig:ite_v_sdr}
\vspace{-3mm}
\end{figure}

We first investigated the feasibility of DEQ-based source separation using the weight-tied network with the architecture described in Fig.~\ref{fig:proposed_method}, which we refer to as WT-UMX hereafter.
Fig.~\ref{fig:ite_v_sdr} shows the relationship between the NFE of WT-UMX and the average SDR.
The average SDR reaches the best SDR score when the NFE, $L$, is set to four, while it gets worse as the NFE increases.
This tendency is similar to previous work in other tasks~\cite{bai2018trellis,bai2019equilibrium,wang2021approximate}.
Although weight-tied networks and DEQ usually require hundreds of NFE to converge, the models that are evaluated are trained with dozens of NFE~\cite{bai2019equilibrium}.
The relationship between NFE and performance, which clearly shows a maximum performance around a specific NFE value, is compatible with another approach \cite{wang2021approximate}, which uses a closed-form representation.
This shows that the best NFE depends on the task and the network architecture. In our case, the model tends to reach the best performance with a relatively small NFE.
This implies that using BLSTM as $f_\theta$ helps the network to utilize all the information of the input sequence early in the iterations, as hypothesized in sec.~\ref{subsec:model}.
We can conclude that it is possible applying weight-tying to UMX and that the required NFE is relatively small comparing to the experiment conducted in other works.

\subsection{Evaluation of the proposed DEQ-UMX}

The evaluation results are shown in Table~\ref{tab:result}.
The number of parameters, the number of multiply-and-accumulate operations (MACs), the SDR values for each instrument, and the average SDR values over all the instruments are compared among several variants of UMX, WT-UMX (from the preliminary study), and our proposed DEQ-UMX.
The MACs do not include the computations for the STFT and MWF.
``UMX large" is a variant of UMX with more layers in the BLSTM block than the original UMX.
``UMX small" is a variant with smaller hidden size (410) than that of UMX (512).

First, we can observe that even if the number of layers of UMX is simply increased, there is almost no improvement in terms of the average SDR.
Also, if we reduce the hidden size of UMX, then the number of parameters will decrease while the average SDR is penalized by 0.09~dB.
WT-UMX with $L=4$ reduces the number of parameters by 30\% and improves the average SDR, with a particular improvement of the SDR for \emph{drums}.
The proposed DEQ-UMX achieved the best average SDR.
We can also see that JFB contributes to improving the SDR values for all instruments with respect to the model trained with implicit differentiation.
In particular, the proposed DEQ-UMX improved the average SDR by 0.29~dB with respect to ``UMX small", which has the comparable number of parameters. 
We also compared the SDR distributions of both methods and observed that the distributions have a similar shape, i.e., there are no extreme outliers for DEQ-UMX but its distribution is shifted to a better SDR score.
These results suggest that the equilibrium point obtained by our proposed method allows for a better separation performance than the ones obtained by the weight-tied network with typical backpropagation and the DEQ-UMX trained by backpropagation with implicit differentiation.
In our proposed method, $L$ used in pre-training was 4 for all instruments, and the number of epochs for pre-training was 400 for \emph{vocals}, 0 for \emph{drums}, and 300 for \emph{bass} and \emph{other}.
After pre-training, $L_{\text{max}}$ was set to 6 for all instruments.
During inference, the NFE always reached $L_{\text{max}}$ in the test set.
Although the MACs of DEQ-UMX are higher than UMX, the number of parameters of DEQ-UMX is lower than for UMX by around 30\%, which enables the implementation of DEQ-UMX in devices with limited memory.

\section{Conclusion}
We proposed a DEQ-based music source separation algorithm called DEQ-UMX.
Inspired by the architecture of UMX, we replaced its BLSTM layers with a DEQ-based BLSTM layer.
The DEQ-UMX trained with JFB achieved higher average SDR than the original UMX on the MSS task, while reducing the number of parameters. 
In future work, we plan to apply the DEQ framework to other state-of-the-art architectures and show its potential for generalization.
We also plan to explore more efficient root-finding schemes to reduce the current computational complexity.

\vfill\pagebreak

\bibliographystyle{IEEEtrans}
\bibliography{strings,refs}

\end{document}